\documentclass[11pt,fleqn]{article}
\usepackage{amsmath,amsfonts,amssymb,latexsym}

\usepackage{epstopdf}

\usepackage{amsfonts,amssymb,cite}
\usepackage{graphicx}

\def\onecol{\onecolumn \mathindent 2em}

\def\noi{\noindent}

\newcommand{\Title}[1]{\noi {{\Large\bf #1}}\\[1ex]}

\newcommand{\Author}[2]{\noi{\bf #1}\\[2ex]\noi{\normalsize\it #2}\\}

\def\au#1{${}^{#1}$}

\newcommand{\Abstract}[1]{\vskip 2mm \begin{center}
        \parbox{16.4cm}{\small\noi #1} \end{center}\medskip}

\def\email#1#2{\footnotetext[#1]{e-mail: #2}\addtocounter{footnote}{1}}

\def\nqq{\hspace*{-2em}}
\def\nhq{\hspace*{-0.5em}}

\def\cm{\hspace*{1cm}}

\def\Jl#1#2{#1 {\bf #2},\ }

\def\ApJ#1 {\Jl{Astroph. J.}{#1}}
\def\CQG#1 {\Jl{Class. Quantum Grav.}{#1}}
\def\DAN#1 {\Jl{Dokl. AN SSSR}{#1}}
\def\GC#1 {\Jl{Grav. Cosmol.}{#1}}
\def\GRG#1 {\Jl{Gen. Rel. Grav.}{#1}}
\def\JETF#1 {\Jl{Zh. Eksp. Teor. Fiz.}{#1}}
\def\JETP#1 {\Jl{Sov. Phys. JETP}{#1}}
\def\JHEP#1 {\Jl{JHEP}{#1}}
\def\JMP#1 {\Jl{J. Math. Phys.}{#1}}
\def\NPB#1 {\Jl{Nucl. Phys. B}{#1}}
\def\NP#1 {\Jl{Nucl. Phys.}{#1}}
\def\PLA#1 {\Jl{Phys. Lett. A}{#1}}
\def\PLB#1 {\Jl{Phys. Lett. B}{#1}}
\def\PRD#1 {\Jl{Phys. Rev. D}{#1}}
\def\PRL#1 {\Jl{Phys. Rev. Lett.}{#1}}

\def\al{&\nhq}
\def\lal{&&\nqq {}}
\def\eq{Eq.\,}
\def\eqs{Eqs.\,}
\def\beq{\begin{equation}}
\def\eeq{\end{equation}}
\def\bear{\begin{eqnarray}}
\def\bearr{\begin{eqnarray} \lal}
\def\ear{\end{eqnarray}}
\def\earn{\nonumber \end{eqnarray}}

\def\nnn{\nonumber\\ \lal }

\def\yy{\\[5pt] {}}

\def\eql{\al =\al}

\def\dst{\displaystyle}
\def\tst{\textstyle}
\def\fracd#1#2{{\dst\frac{#1}{#2}}}
\def\fract#1#2{{\tst\frac{#1}{#2}}}
\def\Half{{\fracd{1}{2}}}
\def\half{{\fract{1}{2}}}

\def\d{\partial}

\def\sign{\mathop{\rm sign}\nolimits}
\def\diag{\mathop{\rm diag}\nolimits}

\def\const{{\rm const}}

\def\then{\ \Rightarrow\ }

\newcommand{\vars}[1]{\left\{\begin{array}{ll}#1\end{array}\right.}

\usepackage{color}

\newcommand {\cG}{\cal G}
\newcommand {\cD}{\cal D}
\newcommand {\G}{\Gamma}

\newcommand {\bp}{\bar \psi}

\def\mN{_\mu^\nu}

\def\R{{\mathbb R}}

\def\kappa{\varkappa}

\def\eqn#1{\eq\eqref{#1}}
\def\rf{\eqref}

\def\sph{spherically symmetric}
\def\ssph{static, spherically symmetric}
\def\Scz{Schwarz\-schild}
\def\bh{black hole}
\def\wh{wormhole}
\def\asflat{asymptotically flat}

\begin{document}
\onecol

\Title{Spinor fields in spherical symmetry. \yy
    Einstein-Dirac and  other space-times}

\Author{K.A. Bronnikov,\au{a,b,c,1} Yu.P. Rybakov,\au{b,2} and Bijan Saha\au{b,d,3}}
    {\small
    \au{a} \ \ Center for Gravitation and Fundamental Metrology, VNIIMS,
                    46 Ozyornaya St., Moscow 119361, Russia;\\
    \au{b} \ \ Peoples' Friendship University of Russia (RUDN University),
                   ul. Miklukho-Maklaya 6., Moscow 117198, Russia;\\
    \au{c}\ \  National Research Nuclear University ``MEPhI''
                    (Moscow Engineering Physics Institute), Moscow, Russia;\\
    \au{d}\ \ Laboratory of Information Technologies, Joint Institute for Nuclear Research, Dubna\\
        \cm     141980 Dubna, Moscow region, Russia
        }

\Abstract
  {We discuss the \ssph\ Einstein-spinor field system in the possible presence of
   various spinor field nonlinearities. We take into account that the spinor field energy-momentum
   tensor (EMT) has in general some off-diagonal components, whose vanishing due to the Einstein
   equations substantially affects the form of the spinor field itself and the space-time geometry.
   In particular, the EMT structure with any spinor field nonlinearities turns out to be the same as that
   of the EMT of a minimally coupled scalar field with a self-interaction potential. Therefore many
   results previously obtained for systems with such scalar fields are directly extended to the
   Einstein-spinor field system. Some special solutions are obtained and discussed, in particular,
   a solution for the Einstein-Dirac system (which lack asymptotic flatness) and some examples
   with spinor field nonlinearities.}

\email 1 {kb20@yandex.ru}
\email 2 {soliton4@mail.ru}
\email 3 {bijan@jinr.ru, homepage http://spinor.bijansaha.ru}

\section{Introduction}

  In the recent past, the spinor description of matter and dark energy was
  used to draw a picture of the Universe evolution within the scope of Bianchi-type
  anisotropic cosmological models
   \cite{1997GRG,2001PRD,2006PRD,2018EChAYa}. It was found that the
  approach in question gives rise to a variety of solutions depending on the choice of
  the spinor field nonlinearity. Owing to its sensitivity to the gravitational field, the
  spinor field brings some unexpected features to the behavior of the gravitational
  fields and cosmological models. Bearing this in mind, in this paper we consider
  nonlinear spinor fields coupled to spherically symmetric gravitational fields.
  Since a variety of astrophysical systems such as stars and black holes are
  fairly well described within spherical symmetry, the use of spinor fields in this
  area might be very promising.

  The existence of off-diagonal components of the energy-momentum tensor (EMT) of
  the spinor field even in the simplest cases together with the Fierz identities relating
  different invariants composed from the bilinear forms of the spinor field
  impose restrictions either on the geometry of space-time or on the behavior of
  the spinor field itself or on both of them. As a result, the spinor approach gives rise to a variety of
  interesting solutions depending on the choice of a spinor field nonlinearity.
  In particular, in cosmology this approach allows us to explain the late-time acceleration
  of the Universe expansion, generates regular solutions and causes rapid isotropization
  of the initially anisotropic Universe. Moreover, the spinor field nonlinearity can
  simulate different types of fluids and some possible kinds of dark energy
  \cite{1997GRG,2001PRD,2006PRD,2018EChAYa}. This success of the spinor
  approach in cosmology leads many authors to consider it in astrophysics and other
  areas as well. For example, a non-Abelian SU(2) Proca field interacting with
  nonlinear scalar and spinor fields were studied in \cite{Dzhun1}. Scattering of a Dirac
  spinor particle in the field of a Schwarzschild black hole was studied in\cite{Cot}.
  The Dirac equation in curved 5D spherically symmetric space-time was studied
  in \cite{Brih}. A nonlinear spinor field minimally coupled to Maxwell and Proca fields
  have been  considered in a spherically symmetric space-time \cite{Dzhun2}.

   In a recent paper \cite{SpinSphere1}, a nonlinear spinor field in spherically
  symmetric space-times was studied, and it was shown that the existence of
  nontrivial EMT components imposes substantial restrictions on both the spinor field
  and the geometry. In the present paper, we make some further observations on
  the properties of the Einstein-spinor field system in \ssph\ space-times and discuss
  some exact solutions. It turns out, in particular, that the EMT of a \sph\ spinor field,
  after eliminating its possible off-diagonal components, has the same structure
  as that of a static scalar field, and therefore a number of results known for the
  Einstein-scalar field system, concerning the possible existence of Killing horizons
  and wormholes, are directly extended to the Einstein-spinor system.
  Next, we discuss exact solutions with the Dirac linear spinor field and some
  kinds of spinor field nonlinearities.

\section{The Einstein-spinor equations}

\subsection{General equations}

  Let us consider a system of (in general, nonlinear) spinor and gravitational fields
  in the framework of general relativity. The action can be written in the form
\beq                     \label{action}
    {\cal S}(g; \psi, \bp) = \int\, \sqrt{-g} d^4x
            \left( \frac{R}{2\kappa} + L_{\rm sp}\right),
\eeq
  where  $R$ is the scalar curvature,  $\kappa = 8 \pi G$, $G$ being the Newtonian
  gravitational constant, and $L_{\rm sp}$ is the spinor field Lagrangian which
  we take in the form
\beq                  \label{lspin}
        L_{\rm sp} = \frac{\imath}{2} \left[\bp \gamma^{\mu} \nabla_{\mu}
    \psi- \nabla_{\mu} \bar \psi \gamma^{\mu} \psi \right] - m \bp \psi - F,
\eeq
  with the spinor field mass $m$ and the nonlinear term $F = F(K)$, where $K$ is one
  of the four expressions $\{I,\,J,\,I+J,\,I-J\}$, and we use the following notations:
\beq                  \label{defs}
           I = S^2, \quad\ J = P^2;\cm
            S= \bar \psi \psi, \quad\  P =i \bar \psi \gamma^5 \psi.
\eeq

      The spinor field equations corresponding to the Lagrangian \eqref{lspin} are
\begin{subequations}            \label{speq}
\bear
    \imath\gamma^\mu \nabla_\mu \psi - m \psi - {\cD} \psi -
             \imath {\cG} \gamma^5 \psi &=&0,           \label{speq1}
\\
    \imath \nabla_\mu \bp \gamma^\mu +  m \bp + {\cD}\bp + \imath {\cG}
            \bp \gamma^5 &=& 0,                                        \label{speq2}
\ear
\end{subequations}
  where $\nabla_\mu \psi = \d_\mu \psi - \G_\mu \psi$ and
  $\nabla_\mu \bp = \d_\mu \bp + \bp \G_\mu$, the matrices $\G_\mu$
  being those of the spinor affine connection. Furthermore, we denote
  ${\cD} = 2 S F_K K_I$ and ${\cG} = 2 P F_K K_J$, with $F_K = dF/dK$,
   $K_I = dK/dI$ and $K_J = dK/dJ$.  Using \eqs \eqref{speq}, it can be shown that
\beq
    L_{\rm sp} = 2 K F_K - F.                      \label{LspinF}
\eeq
  The stress-energy tensor (SET) of the spinor field is given by
\beq                                         \label{SET}
    T\mN = \frac{\imath}{4} g^{\rho\nu}
    \left(\bp \gamma_\mu \nabla_\rho \psi + \bp \gamma_\rho \nabla_\mu \psi
    - \nabla_\mu \bar \psi \gamma_\rho \psi - \nabla_\rho \bp \gamma_\mu \psi \right)
    - \delta\mN  L_{\rm sp}.
\eeq

\subsection{Static spherical symmetry}

  In what follows we consider the general \ssph\ metric
 \beq                                              \label{ds}
        ds^2 = e^{2 \gamma} dt^2 - e^{2 \alpha} du^2 - e^{2 \beta}
                                (d\theta^2 + \sin^2 \theta\, d \varphi^2)
 \eeq
   where $\alpha, \beta, \gamma$ are functions of an arbitrarily chosen radial
   coordinate $u$. The SET $T\mN$ of the spinor field then has, in general, the following
   nonzero components  \cite{SpinSphere1}:
\begin{subequations}        \label{SET}
\bear
        T^0_0 \eql  T_2^2  = T_3^3 = F - 2 K F_K, \label{T00}
\yy
        T^1_1 \eql m S + F,                      \label{T11}
\\
        T^0_1 \eql \frac{1}{4}\cot \theta\, e^{\alpha - \gamma -\beta} A^3,                     \label{T01}
\\
    T^0_2 \eql -\frac{1}{4} (\gamma' - \beta')\,e^{\beta -\alpha - \gamma} A^3,      \label{T02}
\\
    T^0_3 \eql \frac{1}{4} (\gamma' - \beta') \, e^{ \beta -\alpha - \gamma} \sin \theta A^2
    +\frac{1}{4} e^{-\gamma} \cos \theta  A^1.                                  \label{T03}
\ear
\end{subequations}
  where the prime stands for $d/du$, and $A^\mu$  are components of the pseudovector
  $A^\mu= \bp \gamma^5 \gamma^\mu \psi$.
  Since the Einstein tensor $G\mN$ for the metric \rf{ds} is diagonal, due to the Einstein equations
  we must have the off-diagonal components  $T^0_i =0$, whence
\beq                    \label{A1A3}
        A^3 =0, \cm  A^1 = A^2 (\beta'-\gamma') e^{\beta-\alpha} \tan \theta.
\eeq
  These relations impose certain constraints on the components of the spinor field but do not
  restrict the very existence of solutions to the spinor equations and do not directly affect
  the form of the Einstein equations, see more details in \cite{SpinSphere1}.

  It should also be noted that the expressions for $T^0_i$ do not depend on the spinor field
  nonlinearity.

  With \rf{SET} and \rf{A1A3}, the nontrivial components of the Einstein equations can be written as
\begin{subequations}                                   \label{EET}
\bear                           \label{EE00}
    \gamma'' + \gamma' (-\alpha'+2\beta'+\gamma')
                    \eql \kappa e^{2\alpha} (-F + KF_K - mS/2),
\yy                                    \label{EE11}
     e^{2 (\alpha - \beta)}  - 2  \beta' \gamma'  - \beta'{}^2
                    \eql  \kappa e^{2 \alpha} (m S + F),
\yy              \label{EE22}
    - e^{2 (\alpha - \beta)} + \beta'' + \beta' (-\alpha'+2\beta'+\gamma')
                    \eql \kappa e^{2\alpha} (-F + KF_K - mS/2),
\ear
\end{subequations}
   where the first-order equation \rf{EE11} is $G^1_1 = -\kappa T^1_1$ while the other two are
   components of the equations $R\mN = - \kappa (T\mN - \half \delta\mN T)$.
   The conservation law $\nabla_\nu T\mN =0$ leads to
\beq                    \label{cons}
            (m S + F)' + ( \gamma' + 2 \beta')  (m S + 2 K F_K) = 0.
\eeq

  One can notice that due to \eqs \rf{EET} the opportunity $\beta'-\gamma' =0$ (hence $A^1 =0$) that
  could be significant in \rf {A1A3}, should be discarded since in this case the difference of \rf{EE00}
  and \rf{EE22} leads to the impossible equality $e^{2\alpha-2\beta} =0$.

\medskip
   Let us now consider two different choices of the nature of the spinor field. First, suppose
   $K = I = S^2$, then, since $F' = 2 S F_K S'$, from \eqref{cons} we find
   (provided that $m + 2S F_K \ne 0$)
\beq                                               \label{cons1}
            S = S_1  e^{-(\gamma + 2\beta)}, \qquad    S_1 = \const >0.
\eeq
   We take $S_1 > 0$ since the quantity $S = {\overline\psi}\psi$ is positive-definite.

   Second, suppose that $K$ is any of the variants $\{I,\,J,\,I+J,\,I-J\}$, and consider a
   massless spinor field ($m=0$), as was done in cosmology \cite{1997GRG, 2001PRD}.
  Then, assuming $F(K) \ne \const$, \eqn{cons} leads to
\beq                        \label{cons2}
        K = K_1 e^{-2(\gamma + 2\beta)}, \qquad    K_1 = \const.
\eeq
  Evidently, the case $m=0,\ K=I = S^2$ belongs to both two variants, and then $K_1 = S_1^2$.

  Equations \rf{SET}--\rf{cons2} immediately lead to a number of important consequences.

\medskip\noi
{\bf 1.}   By \eqref{T00}, we have $T^0_0 = T_2^2 = T_3^3$. This property is the same
   as is known for  minimally coupled scalar fields with arbitrary self-interaction potentials,
   and this immediately leads to the same conclusion as was proved in \cite{kb-scalar}
   on the basis of this equality, concerning the possible global structure of any
   space-time whose metric is obtained with this SET. According to the {\it global
   structure theorem} from \cite{kb-scalar},

\medskip\noi
   {\sl There can be at most two horizons at which $e^\gamma =0$. Around a static region, horizons
   can only be simple (non-extremal).}
\medskip

  In particular, if there is a static spatial infinity (be it flat, AdS or any other), there can be only one
  simple horizon, similar to that in a \Scz\  \bh.

\medskip\noi
{\bf 2.}  In the case of a linear spinor field when $F = 0$, we have
  $T^0_0 = T_2^2 = T_3^3 = 0$, and the only nontrivial SET component is $T_1^1 = m S$
  which is nonzero only for a massive spinor field. A massless (neutrino) field is ``stealth''
   in the sense that its SET is completely zero.

  Furthermore, a space-time with  $T_1^1 = m S$ cannot be \asflat. Indeed, asymptotic
  flatness requires that at large $r \equiv e^\beta$ the metric should be approximately
  \Scz, that is,
\[
        e^{2\gamma} = 1 - 2M/r + O(r^{-2}),   \cm       e^{2\alpha} = 1 + 2M/r +O(r^{-2}),
\]
  from which it follows that all $G\mN = O(r^{-4})$, and by the Einstein equations the same
  is required for $T\mN$,  whereas by \rf{cons1} $T_1^1 = O(r^{-2})$.

  This conclusion is also true with any nonlinearity $F(I)$, except for the special case where
  this nonlinearity behaves precisely as $-mS$ at small $S$ and thus exactly eliminates
  the mass term $mS$ in $T^1_1$; the other SET components then also behave as
  $o(S)$ at small $S$.

  If $m=0$, then, according to \rf{cons1} and \rf{cons2}, the space-time can be \asflat\ if
  $F(K) \sim K$ or $F(K) = o(K)$ at small $K$.

\medskip\noi
{\bf 3.} A necessary condition for the existence of \wh\ throats is that $T^0_0 - T^1_1 < 0$,
  which  violates the Null Energy Condition. With the tensor \rf{SET} this inequality reads
\beq                    \label{-NEC}
        - mS -2K F_K < 0.
\eeq
  The inequality \rf{-NEC} holds both for a linear massive field with $m>0$ and any
  nonlinearities such that $K F_K > 0$.

\section{Einstein-Dirac solutions}

  Let us consider a linear Dirac spinor field, in which case
  the only nonzero component of the energy-momentum tensor (EMT) $T\mN$ is
  $T^1_1 = mS(x)$, and the conservation law implies $S(x) \sim e^{-(2\beta+\gamma)}$
  (see \eqn{cons1}).

  Let us choose the curvature radial coordinate $u = r$ in the metric \rf{ds} (so that $e^\beta = r$),
  Then, in the general case, the temporal component of the Einstein equations can be
  presented in the integral form:
\beq                      \label{EE0}
        e^{-2\alpha} = 1 - \frac{\kappa}{r} \int T^0_0 r^2 dr - \frac 13 \Lambda r^2,
\eeq
  where $\Lambda$ is the cosmological constant included in this case for generality.
  With $T^0_0 =0$, it follows, in full similarity with the \Scz-de Sitter solution,
\beq                              \label{alf}
        e^{-2\alpha} = 1 - \frac{2M}{r} - \frac 13 \Lambda r^2, \qquad M = \const.
\eeq
  Let us substitute it into the ${1\choose 1}$ component of the Einstein equations,
  having the form
\beq                \label{EE1D}
    e^{-2\alpha}\bigg(\frac{2\gamma'}{r}+ \frac 1{r^2}\bigg) - \frac 1 {r^2}
                     + \Lambda = - \kappa m S.
\eeq
  With \rf{cons1}  we obtain
\beq            \label{EE1}
    1 + 2r \gamma' + (\Lambda r^2 -1)e^{2\alpha} + S_2 e^{2\alpha - \gamma} =0,
\eeq
  where $\alpha(r)$ should be substituted from \rf{alf} and $S_2 = \kappa m S_1 > 0$.
  Equation \rf{EE1} is a linear first-order equation with respect to $y(r)= e^\gamma$:
\beq            \label{EE1a}
    2r y' + \big[1 + (\Lambda r^2 -1)e^{2\alpha}\big] y = -S_2 e^{2\alpha}.
\eeq
  In the general case $\Lambda \ne 0$, $M \ne 0$ its solution can be found in
  quadratures by standard methods, but here we will restrict ourselves to simplest
  special cases.

 \medskip\noi
{\bf 1.} $M = \Lambda =0$. Then from \rf{EE1a} it follows
\beq                                       \label{gam00}
        e^\gamma = \Half S_2 \ln \frac {r_0}{r}, \cm r_0 = \const >0,
\eeq
  and the range of $r$ is $0 < r < r_0$. The metric has the form
\beq
        ds^2 = \frac 14 S_2^2 \Big(\ln \frac {r_0}{r}\Big)^2 dt^2 - dr^2 - r^2 d\Omega^2.
\eeq
  As $r\to 0$ we have $e^\gamma \to \infty$, a repulsive singularity. At $r=r_0$, on the contrary,
  $e^\gamma \to 0$, it is an attracting singularity.

\medskip\noi
{\bf 2.} $M = 0, \ \Lambda \ne 0$. Then \eqn{EE1a} is solved to give
\beq                                        \label{gam0}
        e^\gamma = \frac{S_2}{2} \bigg[-1 + \frac{\sqrt{b^2 - \eta r^2}}{b}
                    \ln \frac {C (b + \sqrt{b^2 - \eta r^2})}{r} \biggr]
                = \frac{S_2}{2} \bigg[-1 + \sqrt{1 - \eta z^2}
                    \ln \Big(C\frac {1 + \sqrt{1 - \eta z^2}}{z} \Big) \bigg],
\eeq
  where $C > 0$ is an integration constant, and we are using the notations
\[
        b = \sqrt{|\Lambda/3|}, \qquad \eta = \sign\Lambda, \qquad z = r/b.
\]

   It is easy to verify that at $r = 0$ in all cases we have a repulsive singularity due to
   $e^\gamma \to \infty$. However, other properties of the metric crucially depend on
   the sign of $\Lambda$.

   If $\Lambda>0$, that is, $\eta =1$, the solution is defined in the range $0< z <z_0 <1$,
   or equivalently $0 < r < r_0 < b$ (where $r_0 = bz_0$), in which the quantity
   $y(r) = e^{\gamma(r)}$ is positive. By definition $r_0$ is the value of $r$ at which
   $y =0$, and since $y'(r_0)$ is finite, the derivative $\gamma'(r) = y'/y$ is infinite at $r=r_0$,
   which leads to a curvature singularity.\footnote
        {This singularity is related to an infinite value of the curvature invariants that
         involve the squared component of the Riemann tensor \cite{BR-book}
         $R^{02}{}_{02} = -e^{-2\alpha} \beta'\gamma'$, where
         the quantities $\alpha$ given by \rf{alf} and $\beta'=1/r$ are finite at $r=r_0$
         while $\gamma' = \infty$.
         }
   Thus the solution exists between two singularities and does not reach the value
   $r = b$ at which $e^{2\alpha}$ would change its sign similarly to the de Sitter metric.
   The value of $z_0 = r_0/b$ does not depend on $b$ but depends on $C$ as shown in
   Fig.\,1a.

\begin{figure}[ht]
\centering
\includegraphics[width=6.5cm]{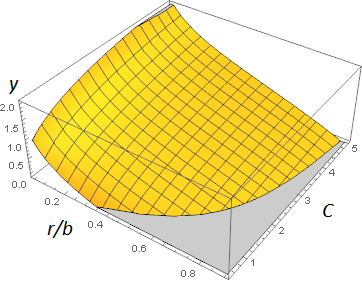}\cm
\includegraphics[width=6.5cm]{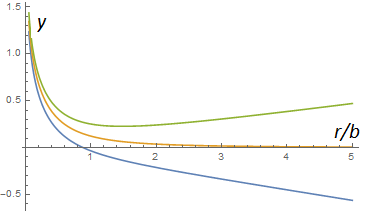}\\
     {\bf a} \hspace{7.5cm} {\bf b}
\caption{\small
        Solutions $y(r) = e^\gamma$ to \eqn{EE1a} for $M=0$: {\bf a} --- 3D plot for
        $\Lambda >0$, $S_2 =1$; {\bf b} --- plots for $\Lambda < 0$, $S_2=1$ and
        $C = 0.8,\ 1,\ 1.2$ (bottom-up).}
\end{figure}

   If $\Lambda < 0$, that is, $\eta  = -1$, the function $e^{-2\alpha} = 1+r^2/b^2$ is the same
   as in the AdS metric and is positive at all $r>0$, but $e^\gamma$ is quite different from its
   AdS counterpart, its behavior depending on the constant $C$. More specifically, if $C < 1$,
   $e^\gamma$ turns to zero at some finite $r=r_0$ and leads to a singularity in the same
   way as in the case $\Lambda >0$. If $C =1$, then $e^\gamma$ remains positive at
   all $r$ but vanishes as $r \to \infty$. Lastly, if $C>1$, the solution is also defined ar
   all $r>0$, and at large $z$ there is a linear asymptotic growth,
    $e^\gamma \approx r (S_2/2) \ln C$, so that $e^{2\gamma} \sim r^2$ as in the AdS metric,
    but in general the AdS relation $\alpha+\gamma =0$ does not hold even asymptotically.

\medskip\noi
{\bf 3.} $\Lambda =0,\ M \ne 0$.  Then we can rewrite \eqn{EE1a} in the form
\beq                \label{EE1b}
        2r (r-2M) y' - 2M y + rS_2 =0,
\eeq
  and its solution is
\beq
        y(r) \equiv e^\gamma =
            + S_2 \bigg[1 -  \frac{\sqrt{r-2M}}{\sqrt{r}} \ln \frac{\sqrt{r}+\sqrt{r-2M}}{C_1}\bigg],
            \qquad C_1 = \const >0.
\eeq
  Note that $M$ is here, in general, not a mass, and there is no reason to assume its
  particular sign.

  In all cases there is a value $r_0$ of the radial coordinate such that $e^\gamma(r_0)
   = y(r_0) =0$ but  $y'(r_0) \ne 0$. It is a singularity (see footnote 1) that separates
   two ranges of $r$. The range $r > r_0$ extends to infinity with the asymptotic
   behavior $e^\gamma \sim \ln(r/r_0)$.

  If $M >0$, the two ranges are $2M < r < r_0$ and $r > r_0$. It is of interest that the
  algebraic curvature invariants are finite at $r = 2M$ but the solution cannot be extended
  beyond this value of $r$ due to loss of analyticity.

  If $M< 0$, the two ranges are $0 < r < r_0$ and $r > r_0$, and the center $r=0$.is a
  repulsive singularity.

\section{Some solutions with nonlinear spinor fields}

  In this section we will describe some solvable examples with massless ($m=0$) nonlinear
  spinor fields.

\medskip\noi
{\bf 1.} Consider first the case of a linear dependence of $F$ on any of the spinor invariants $K$,
  that is, $F=\lambda K$, $\lambda = \const$. In this case the SET has the form
\beq              \label{SET-F}
      T\mN = -\lambda K \diag (1, -1, 1, 1),
\eeq
  whose structure coincides with that known for a massless, minimally coupled
  scalar field, which is canonical if $\lambda K < 0$ and phantom if $\lambda K > 0$.
  Consequently, the Einstein equations lead to the same metrics, the Fisher and anti-Fisher
  ones for the canonical and phantom fields, respectively. A brief unified presentation of
  these metrics involving their all four branches uses the harmonic radial coordinate $u$
  defined by the coordinate condition $\alpha  = 2\beta + \gamma$ \cite{kb73}:
\bearr                   \label{ds-F}
             ds^2 = e^{-2hu}dt^2 - \frac{e^{2hu}}{s^2(k,u)}\bigg[\frac {du^2}{s^2(k,u)} + d\Omega^2 \bigg],
\nnn                     \label{phi}
    s(k,u) := \vars {
                        k^{-1}\sinh ku,  \ & k > 0 \\
                                    u,  \ & k = 0 \\
                        k^{-1}\sin ku,   \ & k < 0,  }
\ear
  where the constants  $h$ and $k$ are also involved in the relation that follows from \eq {EE11}
  with \rf{cons2}
\beq                            \label{int-F}
        k^2 \sign k = h^2 - \kappa\lambda K_1.
\eeq
  If $\lambda K_1 < 0$, we have $k>0$, we are dealing with Fisher's metric \rf{ds-F},
  in which $u \in \R_+$, the value $u =0$ corresponds to spatial infinity where the metric is \asflat,
  and the \Scz\ mass $M$ is equal to $h$. As $u\to \infty$, there is a naked, attracting (if $M>0$)
  singularity with $g_{tt} = e^{2\gamma} \to 0$.

  If  $\lambda K_1 > 0$, the constant $k$ may be zero, positive or negative, and accordingly the
  metric \rf{ds-F} (often called the anti-Fisher metric) splits into three branches.
  In all of them the metric is \asflat\ (again at $u=0$ with the \Scz\ mass $M = h$), but now there
  is no center (the radius $r = e^\beta$ never turns to zero), instead, there are throats (i.e., regular
  minima of the function $r(u)$), and the branch $k < 0$ describes twice \asflat\ wormholes.
  The second spatial infinity corresponds to $u = \pi/|k|$. The (anti-)Fisher metrics have been
  described and discussed in detail in many papers, we will not do it here and refer the reader
  to the papers \cite{BR-book,kb73,kb11,h_ellis} and references therein.

  According to the definitions of the invariants $I, J, K$ (see \rf{defs}), both $I$ and $J$ are
  positive-definite, therefore, if $K$ equals $I$, $J$ or $I+J$, the sign of the energy density
  $T^0_0$ in \rf{SET-F} (in other words, the canonical or phantom nature of the nonlinear
  spinor field) is determined by the coupling constant $\lambda$, and only if $K = I - J$, this
  sign is determined by that of the combination $\lambda K_1$.

\medskip\noi
{\bf 2.}  In the general case, due to the equality $T^0_0 = T^2_2 \then  R^0_0 = R^2_2$,
  the corresponding combination of the Einstein equations admits integration in terms
  of the quasiglobal coordinate $x$ defined by the condition $\alpha + \gamma =0$.
  Denoting $e^{2\gamma} = A(x)$ and $e^\beta = r$, we write the metric in the form
\beq                          \label{ds-Q}
        ds^2 = A(x) dt^2 - \frac{dx^2}{A(x)} - r^2(x) d\Omega^2.
\eeq
  The equation $R^0_0 = R^2_2$ then reads
\beq
        A(r^2)'' - A'' r^2 = 2
\eeq
  (the prime denotes $d/dx$) and is easily integrated giving
\beq                            \label{A'}
        \bigg(\frac{A}{r^2}\bigg)' = \frac{6M - 2x} {r^4}, \cm  A(x) = r^2 \int \frac{6M-2x}{r^4}dx,
        \cm M = \const,
\eeq
  which yields $A(x)$ if $r(x)$ is known.

  Equation \rf{A'} thus makes it possible to find solutions to our problem using the
  inverse problem method: given $r(x)$ in a form of interest for some reasons, from
  \rf{A'} we find $A(x)$, so that the metric is known completely, and the spinor filed
  nonlinearity for which it is a solution is then found from the remaining Einstein equations,
  for example, from the ${1 \choose 1}$ equation having the form
\beq                                                \label{G11}
        \kappa F(K) = \frac 1 {r^2} (1 - A' rr'  - A r'{}^2),       .
\eeq
  while $K$ is already known from \rf{cons2} as $K = K_1/(A r^4)$.

 The same method was used for finding solutions with scalar fields in \cite{pha1,pha2,pha3,pha4}
 and others, where the scalar field and its self-interaction potential were calculated from
 the metric. In the present case such a quantity to be calculated are the spinor field and
 its nonlinearity function.

 \medskip\noi
 {\bf Example:} Let us assume, as in \cite{pha1,pha2,pha3},
\beq                                       \label{r-pha}
        r(x) = \sqrt {b^2 + x^2}\cm  b= \const > 0,
\eeq
  which, as we know from the cited papers, leads to a number of wormhole and black-universe
  solutions. The corresponding function $A(x)$ reads
\bearr                                                       \label{A}
       A(x) =  1 + \frac{C r^2}{b^2} + \frac{3M}{b^3}
                \bigg( bx + r^2 \arctan \frac x b\bigg), \cm C = \const.
\ear
  The metric thus depends on three constants: the ``input'' constant $b$ determining the
  length scale, and two integration constants $M$, equal to the \Scz\ mass if the metric
  is \asflat\ as $x\to\infty$, and $C$ that affects the global properties of the metric.

  In particular, in the case $M=0$ we have $A(x) = 1 + Cr^2/b^2$, so that the solution is
  symmetric with respect to the sphere of minimum radius ($x=0$, $r=b$), is twice
  \asflat\ if $C=0$, de Sitter if $C < 0$ and AdS if $C > 0$. From  \rf{G11} we then find
\beq
        \kappa F(K) = \frac {b^2}{r^4} - \frac {3Cx^2}{b^2 r^2}.
\eeq
  With $K(x)$ equal to $K_1/(A r^4)$, it follows that $F(K)$ can only be found in a parametric
  form. Only the case $C=0$, when the solution describes the Ellis twice \asflat\ wormhole
  \cite{kb73, h_ellis}, is simple enough: we then have $F = \const\cdot K$, in agreement with
  the fact that the Ellis wormhole is described by a special case of the anti-Fisher solution.

  In the general case of \eqn{A} we can evidently also obtain $F(K)$ in a parametric form.
  In all such cases an important question concerns the monotonicity ranges of both $K$ and
  $F$ as functions of $x$, which should be considered for each special solution.

\section{Conclusion}

  We have studied the possible properties of \ssph\ space-times in general relativity with a
  source in the form of linear or nonlinear spinor fields. It turns out that if we exclude
  the ``pathological'' spinor field structures leading to the emergence of off-diagonal EMT
  components (which is necessary due to the Einstein equations), then the algebraic
  structure of the spinor EMT completely coincides with that of the EMT of minimally coupled
  scalar fields with arbitrary self-interaction potentials. It means that the whole set of metrics
  satisfying the Einstein-spinor equations is the same as for the Einstein-scalar equations,
  and the arbitrariness in scalar field potentials now has a counterpart in the arbitrariness
  of spinor field nonlinearities.

  A very important issue is the stability of static configurations. It is well known that the same
  space-time geometry may be stable or unstable depending on the properties of its material
  source. For example, the Ellis \wh\ is known to be unstable when considered with a phantom
  scalar field as a source \cite{kb11, hay02, ggs1} but is stable with a source in the form of
  a perfect fluid having some specially chosen equation of state \cite{stab-sha}. Many other
  scalar-vacuum solutions in general relativity have turned out to be unstable, including Fisher's
  solution \cite{kb-hod} and many of the ``black-universe'' solutions with the metric
  \rf{ds-Q}, \rf{r-pha}, \rf{A} \cite{kb-kon}. It would be of great interest to find out whether the
  same geometries can be stable with a spinor source, and we hope to explore this problem,
  among others, in the near future.

\subsection*{Acknowledgments}

  The work of K.B. was partly performed within the framework of the Center FRPP
  supported by MEPhI Academic Excellence Project (contract No. 02.a03.21.0005, 27.08.2013)
  and also partly funded by the RUDN University Program 5-100.

The work of BS was supported in part by a joint Romanian-JINR, Dubna
Research Project, Order no.396/27.05.2019 p-71.

\end{document}